\documentclass[preprint]{aastex631}

\usepackage{multirow}
\usepackage{amsmath}
\usepackage{appendix}

\shorttitle{The origin of CMEs with intricate magnetic structures}
\begin{document}

\title{The Birth of a Major Coronal Mass Ejection with Intricate Magnetic Structure from Multiple Active Regions}

\correspondingauthor{J. H. Guo}
\email{jinhan.guo@nju.edu.cn}

\author[0000-0002-4205-5566]{J. H. Guo}
\affiliation{School of Astronomy and Space Science and Key Laboratory of Modern Astronomy and Astrophysics, Nanjing University, Nanjing 210023, China}

\affiliation{Centre for mathematical Plasma Astrophysics, Department of Mathematics, KU Leuven, Celestijnenlaan 200B, B-3001 Leuven, Belgium}

\author[0000-0002-9908-291X]{Y. W. Ni}
\affiliation{School of Astronomy and Space Science and Key Laboratory of Modern Astronomy and Astrophysics, Nanjing University, Nanjing 210023, China}

\author[0000-0003-3364-9183]{B. Schmieder}
\affiliation{Centre for mathematical Plasma Astrophysics, Department of Mathematics, KU Leuven, Celestijnenlaan 200B, B-3001 Leuven, Belgium}
\affiliation{LIRA, Observatoire de Paris, CNRS, UPMC, Universit\'{e} Paris Diderot, 5 place Jules Janssen, F-92190 Meudon, France}

\author[0000-0002-9293-8439]{Y. Guo}
\affiliation{School of Astronomy and Space Science and Key Laboratory of Modern Astronomy and Astrophysics, Nanjing University, Nanjing 210023, China}
\author[0000-0002-7153-4304]{C. Xia}
\affiliation{School of Physics and Astronomy, Yunnan University, Kunming 650500, China}

\author[0000-0003-0713-0329]{P. Devi}
\affiliation{Department of Physics, DSB Campus, Kumaun University, Nainital 263001, India}

\author[0000-0002-3518-5856]{R. Chandra}
\affiliation{Department of Physics, DSB Campus, Kumaun University, Nainital 263001, India}

\author[0000-0002-1743-0651]{S. Poedts}
\affiliation{Centre for mathematical Plasma Astrophysics, Department of Mathematics, KU Leuven, Celestijnenlaan 200B, B-3001 Leuven, Belgium}
\affiliation{Institute of Physics, University of Maria Curie-Skłodowska, ul.\ Radziszewskiego 10, 20-031 Lublin, Poland}

\author[0000-0003-0020-5754]{R. Joshi}
\affiliation{Rosseland Centre for Solar Physics, University of Oslo, PO Box 1029, Blindern 0315, Oslo, Norway}
\affiliation{Institute of Theoretical Astrophysics, University of Oslo, PO Box 1029, Blindern 0315, Oslo, Norway}

\author[0000-0002-4391-393X]{Y. H. Zhou}
\affiliation{Centre for mathematical Plasma Astrophysics, Department of Mathematics, KU Leuven, Celestijnenlaan 200B, B-3001 Leuven, Belgium}

\author[0000-0001-6024-8399]{H. T. Li}
\affiliation{School of Physical Science and Technology, Southwest Jiaotong University, Chengdu 611756, China}

\author[0000-0002-7289-642X]{P. F. Chen}
\affiliation{School of Astronomy and Space Science and Key Laboratory of Modern Astronomy and Astrophysics, Nanjing University, Nanjing 210023, China}

\begin{abstract}

Coronal mass ejections (CMEs) are the eruptions of magnetised plasma from the Sun and are considered the main driver of adverse space weather events. Hence, undrstanding its formation process, particularly the magnetic topology, is critical for accurate space weather prediction. Here, based on imaging observations and three-dimensional (3D) data-constrained thermodynamic magnetohydrodynamical (MHD) simulation in spherical coordinates, we exhibit the birth of a CME with intricate magnetic structure from multiple active regions (ARs) due to 3D magnetic reconnection. It is observed as a coronal jet between active regions, accompanied by the back-flowing of filament materials along the jet spine after the passage of the eruptive filament. This jet connects two dimming regions within different active regions. This is an observational proxy of 3D magnetic reconnection between the CME flux rope and the null-point magnetic field lines crossing active regions. Hereafter, the thermodynamic data-constrained MHD simulation successfully reproduces the observed jet and the reconnection process that flux ropes partake in, leading to a CME flux rope with a complex magnetic structure distinct from its progenitor. The generality of this scenario is then validated by data-inspired MHD simulations in a simple multipolar magnetic configuration. This work demonstrates the role of multiple active regions in forming CMEs with intricate magnetic structures. On the one hand, a non-coherent flux rope where not all twisted magnetic field lines wind around one common axis is naturally formed. On the other hand, our findings suggest that the topology of a real CME flux rope may not be solely determined by a single active region, particularly during periods of solar maximum.

\end{abstract}


\section{Introduction} \label{sec:intro}

Coronal mass ejections (CMEs) are the most violent explosions in the solar system, releasing vast amounts of magnetised plasma from the Sun into interplanetary space \citep{Gosling1993}. During periods of solar maximum, CMEs can even occur several times a day \citep{Webb1994}. If a CME propagates along the Sun-Earth line, it can significantly disturb the near-Earth space environment, posing risks to advanced human technology \citep{Chen2011, Schmieder2013, Schrijver2015}. In particular, the southward magnetic fields carried by CMEs can cause magnetic reconnection with Earth's intrinsic magnetic fields, leading to geomagnetic storms and auroras. Consequently, diagnosing the magnetic structure of CMEs in observations before their full development is crucial for space weather prediction, yet it remains an open question.

Although the magnetic configuration of CME progenitors (such as filaments) remains elusive \citep{Ouyang2015, Patsourakos2020}, it is well-accepted that the majority of CMEs incorporate a twisted magnetic flux rope (MFR), manifested as the core and part of dark cavity in the typical three-part structures \citep{Chen2011, Song2022, Guo2023}. In addition, in-situ profiles of their interplanetary counterparts measured by satellites often exhibit smooth and large-angle rotation, generally considered evidence of twisted magnetic fields \citep{Burlaga1981}. Consequently, most space weather prediction tools insert an analytical and unstable flux-rope model at the super-Alfv\'enic point to drive CMEs \citep{Verbeke2019, Maharana2021}. However, the analytical flux-rope models are commonly coherent, with the twisted field lines winding around one common axis. In contrast, CME flux ropes may become more intricate in real solar magnetic environments, e.g., composed of open and closed twisted field lines, when propagating from the solar surface to interplanetary space, as exhibited in \citet{Guo2024}. This suggests that the perspective of a coaxial twisting CME flux rope should be questionable.

It is widely accepted that magnetic reconnection plays a pivotal role in the formation and development of CME flux ropes. In a two-dimensional (2D) scenario, known as the CSHKP model \citep{Carmichael1964, Sturrock1966, Hirayama1974, Kopp1976}, magnetic reconnection mainly takes place in the overlying envelope loops, which inject poloidal fluxes into the rising CME flux rope, leading to the growth of flux rope and forming underlying flare loops, while keeping the toroidal flux constant. As such, the magnetic flux reconstructed from some interplanetary magnetic clouds is well comparable to that computed from the regions swept by flare ribbons in the solar source regions \citep{Qiu2007, Hu2014, Wangws2017, Thalmann2023}. In a 2D scenario, the magnetic structures of CMEs maintain a coherent flux rope with one common axis all the time. However, in a more realistic 3D scenario, flux ropes can also partake in magnetic reconnection, causing so many 3D phenomena, such as rotation \citep{Shiota2010, Lugaz2011, Guo2023b}, footpoint drifting \citep{Aulanier2019, Dudik2019, Gou2023} and even failed eruption \citep{Chen2023, Jiang2023}. In this process, CME flux ropes could deviate from their pre-eruptive structure, significantly complicating the prediction of their magnetic structure and potential geomagnetic effects. Therefore, finding the manifestation of such 3D magnetic reconnection during the CME formation and its implications for CME flux ropes in space weather prediction is crucial.

As is well known, anti-parallel magnetic fields typically exist near null points and their associated separatrices. Therefore, 3D magnetic reconnection involving eruptive flux ropes is likely to occur as they pass through these areas of intense magnetic field distortion, as demonstrated in some MHD simulations \citep{Driel-Gesztelyi2014, Jiang2013, Jiang2018}. Furthermore, such null points separating large-scale magnetic systems are prone to form mainly during periods of solar maximum due to multiple active regions \citep{Wangjx2015}. In this paper, we investigate 3D magnetic reconnection resulting from multiple active regions during the formation of a CME at the start of solar cycle 25, utilising imaging observations and data-constrained thermodynamic MHD simulations. We aim to identify observational proxies for this physical process and demonstrate its implication on CME magnetic topology with MHD simulations, thereby enhancing our ability to evaluate the complexity of a CME based on remote-sensing observations. Observational features and simulation results are respectively presented in Sections \ref{sec:obs} and \ref{sec:sim}, followed by conclusions and discussions in Section \ref{sec:sum}.

\section{Observations}\label{sec:obs}

\begin{figure*}
  \includegraphics[width=15cm,clip]{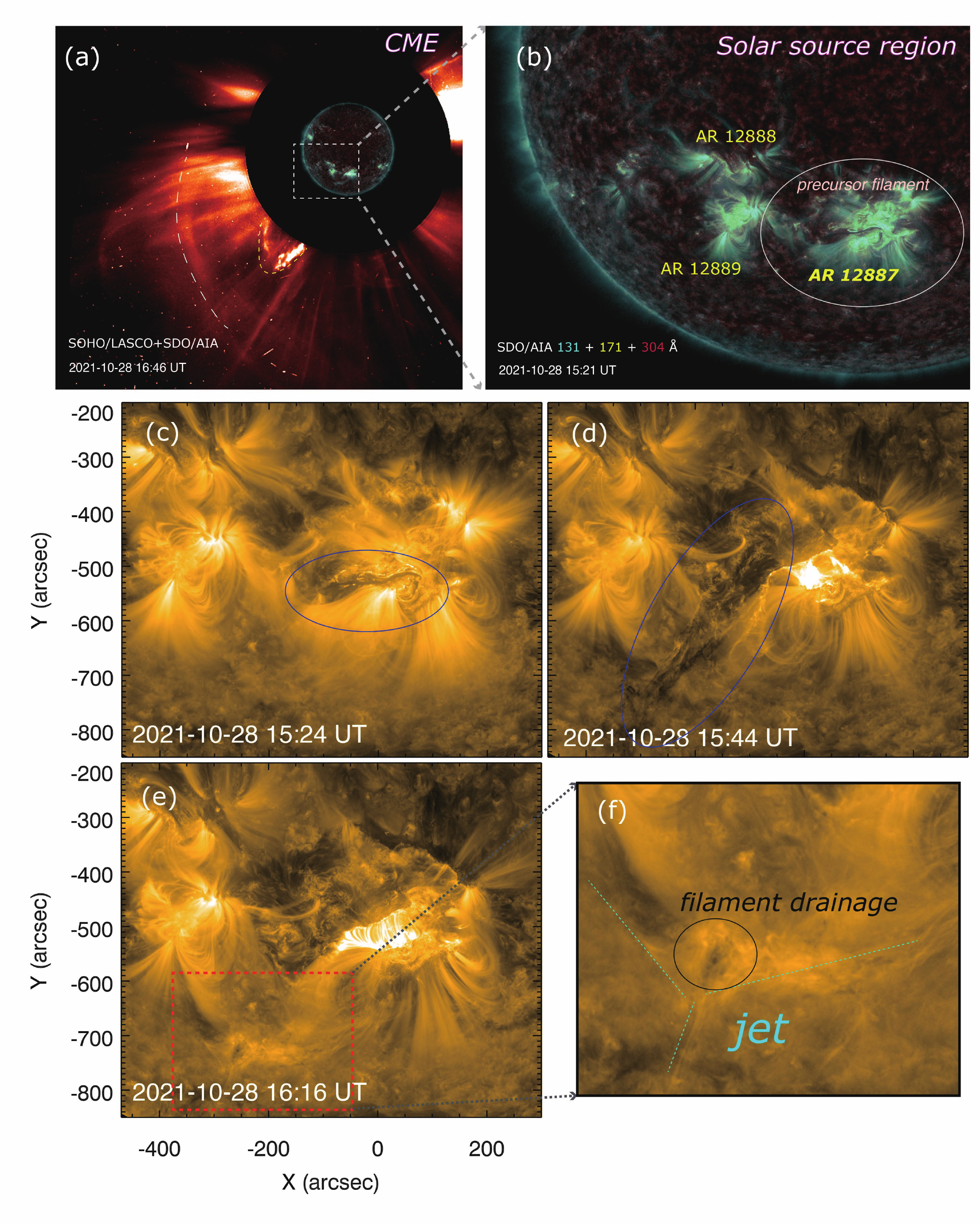}
  \centering
  \caption{Overview of the major CME and its progenitor (filament) evolution in source regions. Panel (a) exhibits the CME captured by SOHO/LASCO C2 white-light coronagraph, wherein the inserted figure shows the composite images in 131, 171 and 304 \AA\ channels of SDO/AIA, zoomed in panel (b). Panels (c)--(e) show the filament eruption in the 171 \AA\ channel of SDO/AIA at 15:24, 15:44 and 16:16 UT, respectively. The blue ellipse and red rectangle represent the eruptive filament and the jet after the filament passage. Panel (f) shows the zoomed figure of the jet in panel (e). }
  \label{figure1}
\end{figure*}

\begin{figure*}
  \includegraphics[width=15cm,clip]{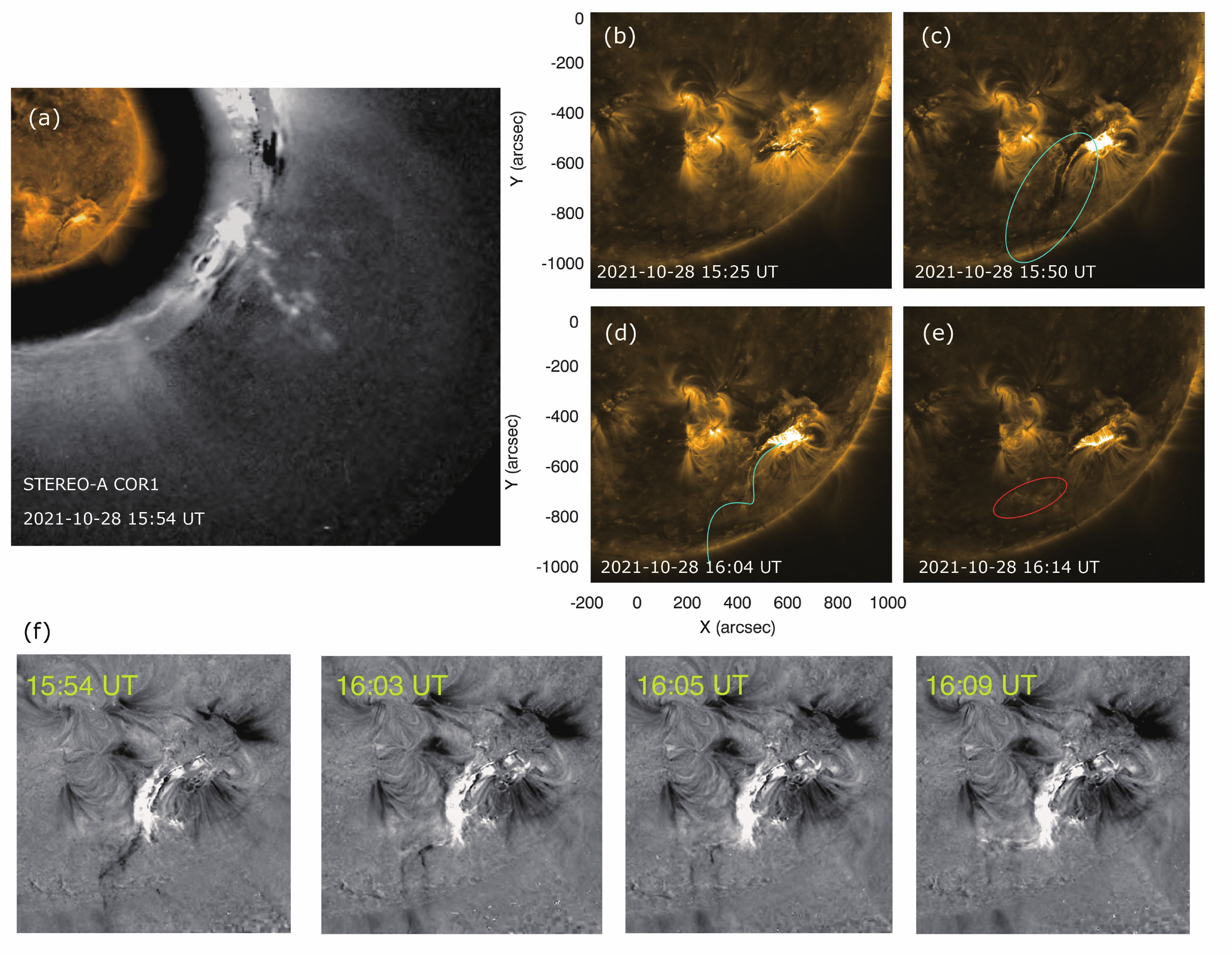}
  \centering
  \caption{STEREO-A observations of the CME event. Panel (a) exhibits the CME captured by STEREO-A COR1 white-light coronagraph, wherein the inserted figure shows the image in 171 \AA\ channel of STEREO-A/EUVI. Panels (b)--(e) show the filament eruption in the 171 \AA\ channel of STEREO-A/EUVI at 15:25, 15:50, 16:04 and 16:14 UT, respectively. The cyan and red ellipses represent the eruptive filament and the brightening after the filament passage. Panel (f) depicts the interaction between the eruptive filament and the null point, as seen in base-difference images at the 171 \AA\ wavelength.}
  \label{figureS1}
\end{figure*}

We focus on the GOES X1.0 flare that peaked at approximately 15:35 UT on 2021 October 28, triggered by a filament eruption in NOAA active region 12887 (designated as the main eruptive active region). This filament eruption also led to a CME featuring a three-part structure in SOHO/LASCO \citep{Brueckner1995} and STEREO-A/COR1 observations \citep{Kaiser2008}: bright core, dark cavity, and bright leading front \citep{Devi2022, Guo2023}, as shown in Figures~\ref{figure1} and \ref{figureS1}. In our previous work \citep{Guo2023}, we have examined the phenomena concentrating on the main eruptive active region, such as the magnetic structure of the filament, reconnection geometries and dynamics of flare ribbons using a data-driven MHD simulation. This paper mainly discusses how these two surrounding active regions influence CME magnetic structure and the corresponding observational phenomena with imaging observations and data-constrained MHD simulations in spherical coordinates.

\begin{figure*}
  \includegraphics[width=15cm,clip]{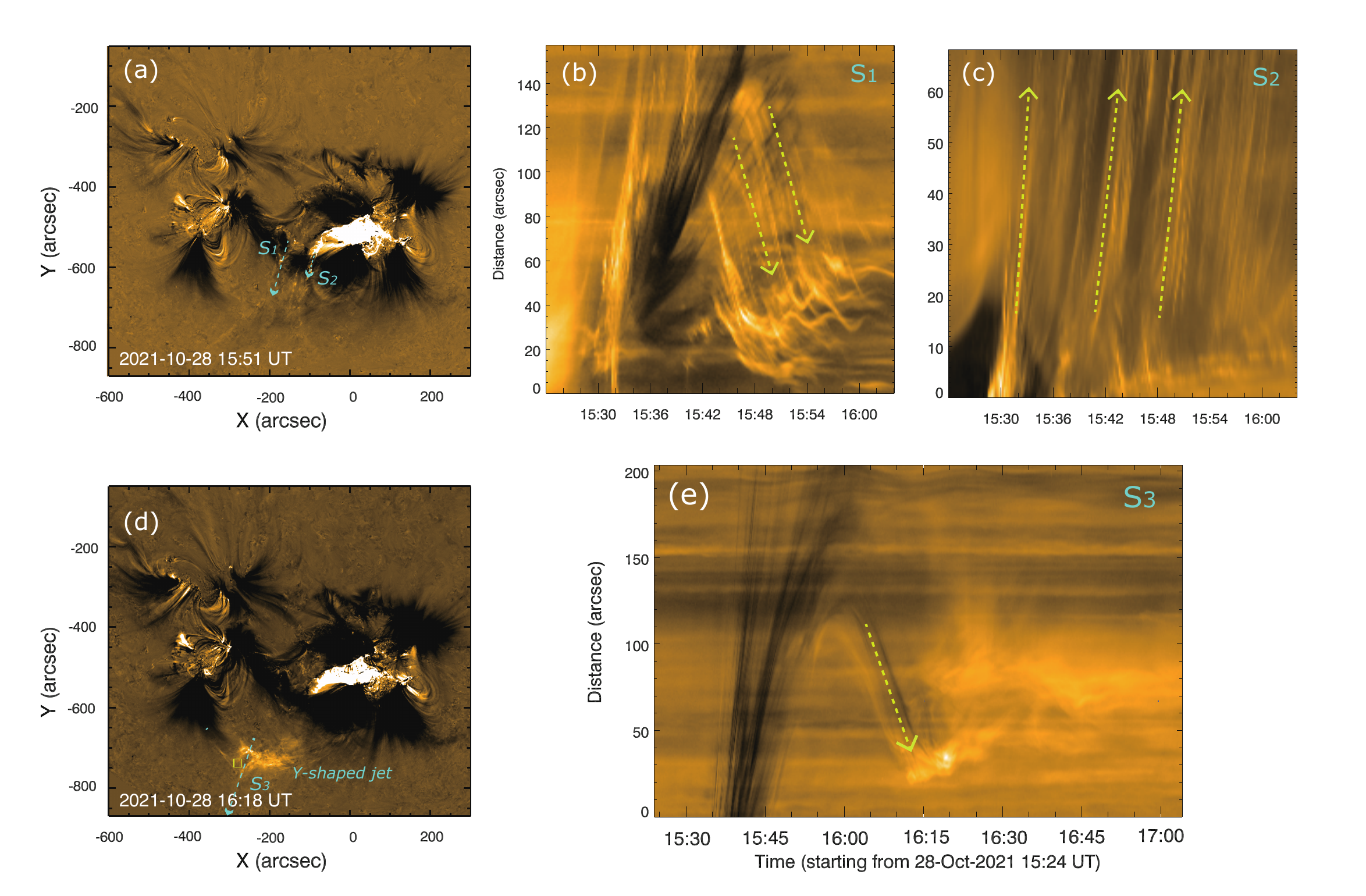}
  \centering
  \caption{Panels (a) and (d) show the base-difference images in the 171 \AA\ channel of SDO/AIA at 15:51 and 16:18 UT, respectively. Panels (b) and (c) show the time-distance diagrams along slices S1 and S2 drawn in panel (a) and in panel (e) along slice S3 drawn in panel (d). The small yellow rectangle in panel (d) represents the area for DEM computation.}
  \label{figure2}
\end{figure*}

Figures~\ref{figure1}c--\ref{figure1}e illustrate the early evolution of the CME captured by SDO/AIA \citep{Pesnell2012} at 171 \AA\  wavelength, from which one can see that a filament leans toward the eastern active regions (Figure~\ref{figure1}d), departing from a radial trajectory. This inclined eruption may be attributed to non-uniform distribution of magnetic pressure (shown in Figure~\ref{figure3}e), as demonstrated in previous works \citep{Chen2000, Panasenco2013, Gui2011, Shen2011, Zhangqm2022, Liuqj2024}. Moreover, some phenomena indicating external reconnection between the eruptive filament and background magnetic fields are also found. For instance, a jet amid active regions is visible after the filament passage (Figures~\ref{figure1}e and \ref{figure1}f), accompanied by certain filament materials flowing back along the jet spine (Figure~\ref{figure2}e). Apart from the SDO observations, this event was also captured by the STEREO-A satellite, located approximately $37^{\circ}.5$ east of the Sun-Earth line \footnote{https://stereo-ssc.nascom.nasa.gov/where.shtml}, providing imaging data from a side-view perspective. As illustrated in Figure~\ref{figureS1}, the CME three-part structures (Figure~\ref{figureS1}a) are clearly visible, along with the eastward drift of filament materials (Figures~\ref{figureS1}b–e) and the backflow of filament materials accompanied by resultant brightening (Figure~\ref{figureS1}f) in the low-corona observation. These features indicate interactions between the eruptive filament and surrounding coronal magnetic fields may exist. In addition to the observations in the low corona, the graduated cylindrical shell (GCS) reconstruction for CME flux rope \citep{Thernisien2011} showed a tilt angle of approximately 61.5$^{\circ}$ concerning the latitudinal direction \citep{Lixl2022}, deviating from the pre-eruptive filament which is nearly aligned with the latitude line. This suggests that specific physical processes contribute to the change in the magnetic structure of the CME flux rope.

To further investigate the 3D magnetic reconnection involving multiple active regions, in Figure~\ref{figure2}, we cut several slices to create time-distance diagrams that illustrate the dynamics. As shown in Figure~\ref{figure2}b, certain dark filament materials accompanied by bright materials are ejected outward at around 15:30 and fell back at around 15:45 UT, accompanied by oscillations of coronal loops between active regions for 2--3 periods. Simultaneously, on its western side, periodic outflows are also observed (Figure~\ref{figure2}c). Subsequently, at around 16:15 UT, a Y-shaped jet between active regions is identified after the filament passage, as shown in Figure~\ref{figure2}d. Moreover, back-flowing is found and accompanied by bright flows along the jet spine, which strikes the jet base and results in localized brightening (Figure~\ref{figure2}e). The mean temperature, derived using the differential emission measure (DEM) method \citep{Su2018}, indicates that the jet base (yellow rectangle in Figure~\ref{figure2}d) is heated to approximately $1.5\;$MK. Furthermore, base-difference images reveal extended coronal dimming regions beyond the main eruptive active region hosting the flare (Figure~\ref{figure2}d), suggesting that the formation of this CME may involve contributions from multiple active regions. 

In summary, based on imaging observations, we find evidence indicating the occurrence of 3D magnetic reconnection during CME formation, driven by interaction among multiple active regions. The key observational phenomena include:

\begin{enumerate}
\item{The axial direction of CME flux rope, reconstructed using GCS reconstruction, deviates from its precursor filament in the solar source region;}
\item{Coronal jets between active regions are observed, with filament materials flowing back along the jet spine;}
\item{Extended coronal dimming regions are detected beyond the main active region hosting the flare.}
\end{enumerate}

To investigate further, we conduct a data-constrained thermodynamic MHD simulation to explore how 3D external magnetic reconnection driven by interactions among multiple active regions influences the magnetic structure of the CME flux rope.

\section{Data-constrained Thermodynamic Semi-Relativistic MHD Modeling in SPHERICAL COORDINATES}\label{sec:sim}

The numerical modelling we adopt is ``Data-constrained thermodynamic semi-relativistic MHD Modelling in Spherical coordinates'', which is implemented in the MPI-AMRVAC framework \citep{Xia2018, Keppens2023}. For comprehensive reviews of data-driven models of solar eruptions, refer to \citet{Jiang2022} and \citet{Guoy2024}. Compared to our previous models \citep{Schmieder2024}, this offers the following advances. First, it incorporates an energy equation that accounts for field-aligned thermal conduction, background heating, and optically thin radiation losses. Second, it applies a semi-relativistic correction \citep{Gombosi2002} to speed up the computations with strong magnetic fields and utilises the constrained transport (CT) on staggered grids \citep{Gardiner2005} to ensure machine-accurate control of magnetic-field divergence. This enables the modeling of strong magnetic fields in observations exceeding 2000~G. Furthermore, this modelling is established in spherical coordinates, allowing us to effectively simulate the impacts of large-scale coronal magnetic fields, such as the interactions between multiple active regions in this paper. For further details on the numerical setup, the reader is referred to Appendix \ref{num}.
\begin{figure}
  \includegraphics[width=15cm,clip]{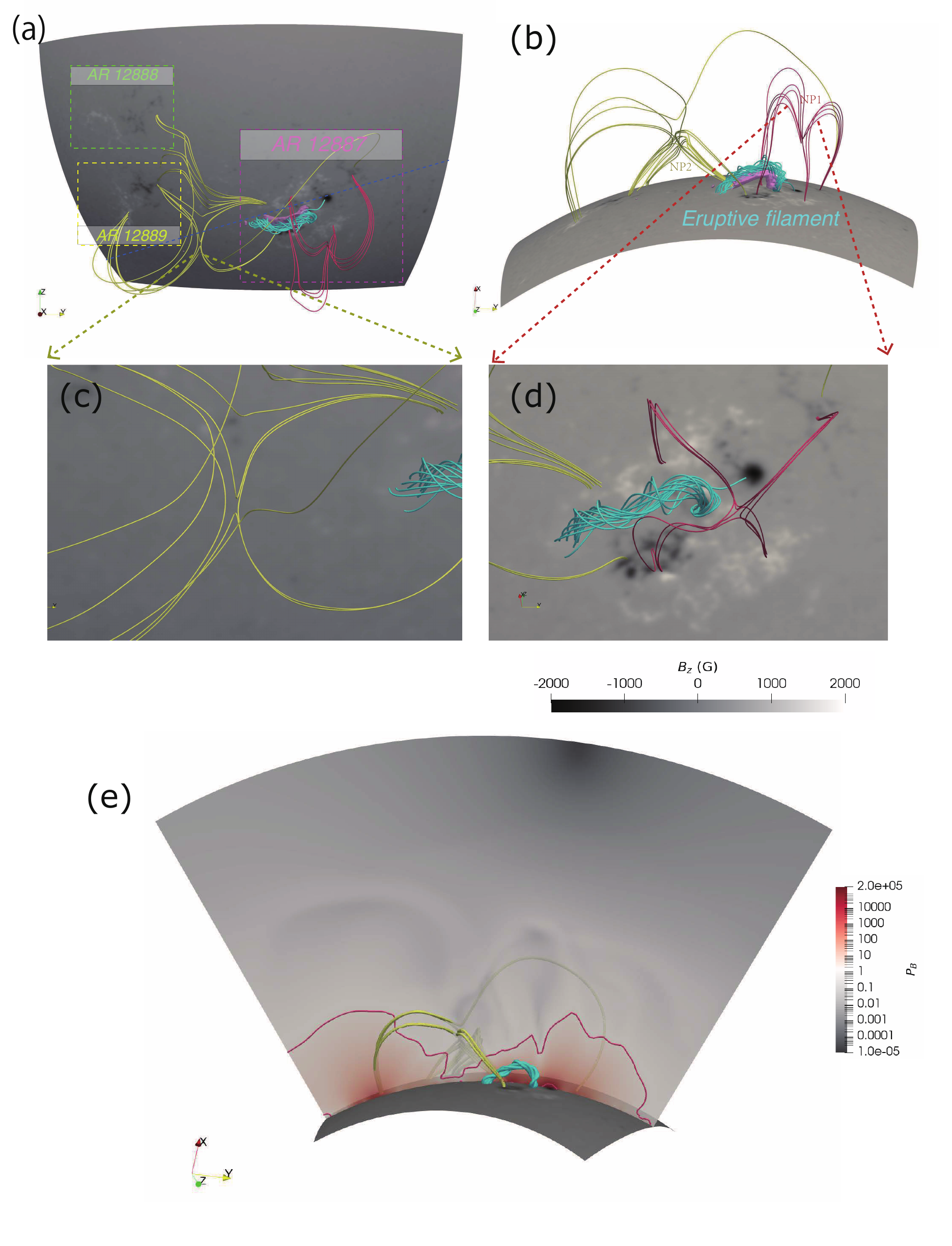}
  \centering
  \caption{Initial 3D magnetic fields after the MHD relaxation. The cyan lines depict the twisted magnetic flux rope to drive the CME, and the semi-transparent contours represent the filament where the number density exceeds $10^{10}$  cm$^{-3}$. The yellow and wine-red lines illustrate the overlying magnetic fields connecting different active regions and those within NOAA AR 12887. Panels (a) and (b) present top and side views of 3D magnetic fields, respectively. Panels (c) and (d) zoom the null-point structure between the active regions and within NOAA AR 12887, respectively. Panel (e) shows the distribution of the magnetic pressure, in which the pink line shows the contour of the value of 2 (normalized unit). The plane is cut off from the blue dotted line in panel (a).} \label{figure3}
\end{figure}

\begin{figure}
  \includegraphics[width=15cm,clip]{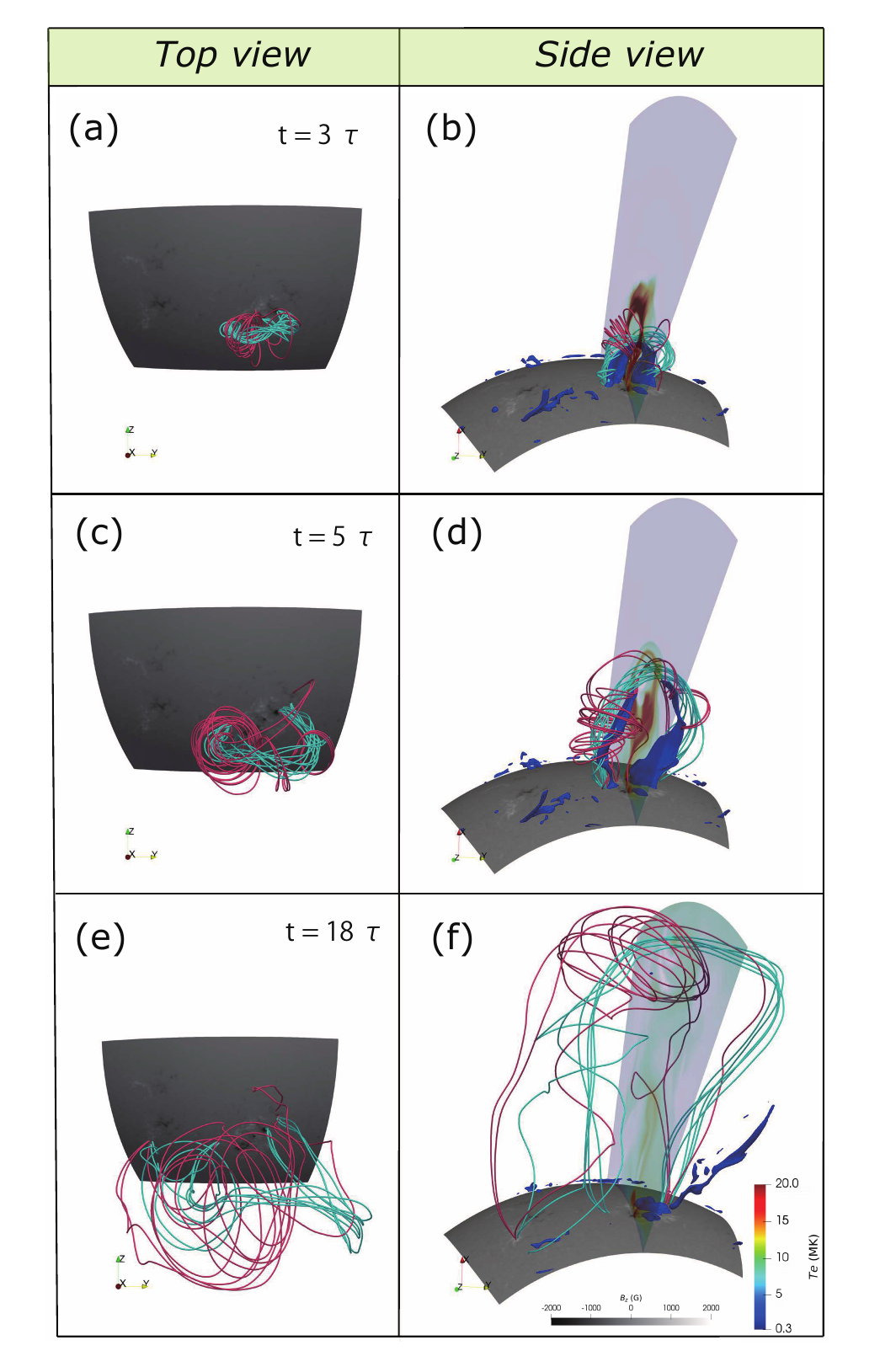}
  \centering
  \caption{Evolution of 3D magnetic fields during the eruption. Panels from the top to bottom display snapshots at 3 $\tau$, 5 $\tau$ and 18 $\tau$, respectively. $\tau$ is defined as 0.01 normalized unit time, corresponding to 1 minute in physical time. The left and right panels show top and side views. The cyan and wind-red lines are traced from the core cold eruptive filament and peripheral hot plasma resulting from magnetic reconnection. The blue semi-transparent contours indicate regions where the number density exceeds $10^{10}\ \rm cm^{-3}$, roughly outlining the eruptive filament.  \label{figure4}}
\end{figure}

\begin{figure}
  \includegraphics[width=15cm,clip]{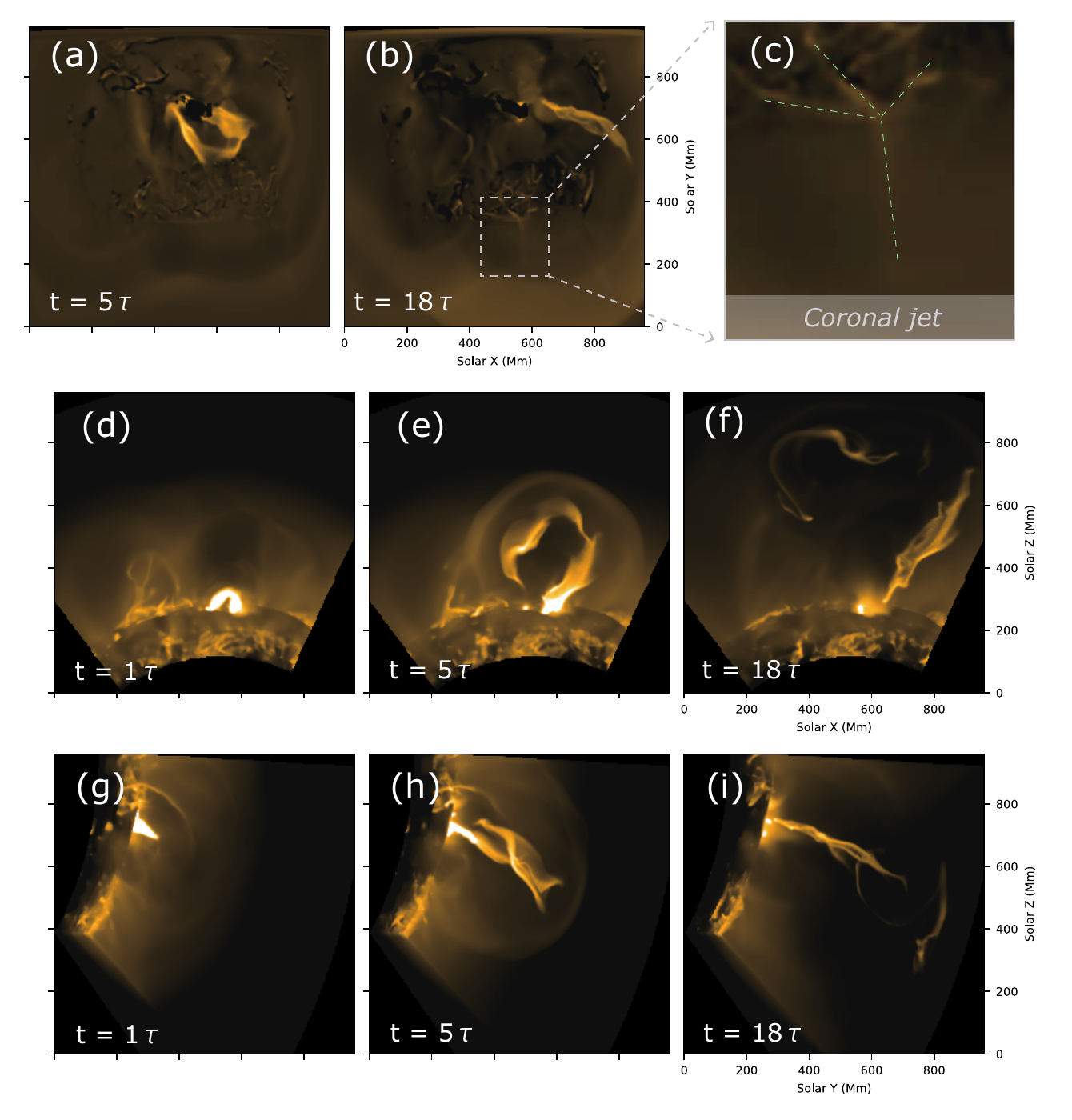}
  \centering
  \caption{Synthesized radiation images in the 171 \AA\ channel of SDO/AIA. Panels (a) and (b) present the base-difference images viewed from the top at 5 $\tau$ and 18 $\tau$, respectively. Panel (c) features a zoomed-in view of the jet from panel (b). Panels (d)--(f)/(g)--(i) display images viewed from the side/end at 1, 5 and 18 $\tau$, respectively. (An animation of the synthesized radiation images is available.) \label{figure5}}
\end{figure}

Figure~\ref{figure3} illustrates the 3D magnetic field configuration at the initial state, from which one can see a twisted magnetic flux rope (cyan lines) and its hosting cold filament materials (pink semi-transparent contours), driving the CME formation. Two X-shaped null-point structures are visible regarding the overlying magnetic fields: one within NOAA AR 12887 due to multi-polarities inside one active region (wine-red lines, referred to NP1), and another connecting three active regions (yellow lines, referred to NP2). As illustrated in simulation results (Figures~\ref{figure4} and \ref{figure5}), the interactions between the CME flux rope and two null-point structures cause the drifting of its two legs (Figure~\ref{figure4}). In observations, NP1 is associated with multiple ribbons and remote dimmings in the main eruptive active region (Figure 1 in \citet{Guo2023} and Figure~\ref{figure2} in this paper), while NP2 should be related to coronal jets and remote dimmings in other active regions (Figure~\ref{figure2}). The co-spatial alignment between the null point found in the simulation and the Y-shaped jet in observations indicates the potential role of this null point in the development of CMEs. To explain the inclined eruption in observations, we present the magnetic-pressure distribution in Figure~\ref{figure3}e. It reveals a high-pressure region in the west, associated with the sunspot, and a low-pressure region in the east, linked to the null point. This pressure asymmetry drives the eastward drift of the filament, as observed in Figures~\ref{figure1} and \ref{figureS1}, consistent with findings from previous studies \citep{Gui2011, Shen2011, Zhangqm2022, Liuqj2024}.

Figure~\ref{figure4} shows the 3D evolution of flux-rope magnetic fields during the eruption, highlighting several physical processes predicted by the standard 2D flare model. For instance, the eruptive flux rope consists of a cold core representing the filament material surrounded by hot plasma formed through magnetic reconnection. The peripheral field lines filled with hot plasma are more twisted than the inner pre-existing cold core (Figures~ref{figure4}c and d), leading to a CME with a weakly twisted core and strongly twisted outer shell \citep{Guo2023}. Additionally, such data-constrained MHD simulation also reproduces certain 3D phenomena in observations. First, it is seen that filament material drains along the flux-rope field lines. Second, certain flux-rope field lines extend into nearby active regions, leading to the migration of CME flux-rope footpoints and the change of its axis orientation. This indicates that the 3D magnetic reconnection that CME flux ropes take part in should occur.

To directly compare our simulation results with observations, we synthesise EUV radiation images at the EUV wavelengths (the method refers to Appendix \ref{radia}), as shown in Figure~\ref{figure5}. The top panels display base-difference synthesised images viewed from the top, which replicates some typical observational features, e.g., an eruptive filament (Figure~\ref{figure5}a), extended dimmings (Figure~\ref{figure5}b) and a Y-shaped jet between active regions (Figure~\ref{figure5}c). The middle panels present side-view images that illustrate the evolution of the eruptive filament and its interaction with the nearby fan-spine structure. In particular, at $t=18~\tau$, we identify a bending at the endpoint of the eruptive filament (also near the fan-spine structure), serving as evidence of magnetic reconnection in which the CME flux rope takes part. The bottom panels of Figure~\ref{figure5} display the CME three-part structure in end-view synthesized radiation images, which closely resemble the morphology observed by STEREO-A COR1 (Figure~\ref{figureS1}a). However, it should be noted that our simulation domain does not fully encompass the field of view of STEREO-A COR1 (1.3–4 solar radii). In a word, this simulation reproduces some fundamental features of the CME in observations, providing a valuable opportunity to investigate the 3D evolution of coronal magnetic fields underlying these phenomena.

\begin{figure}
  \includegraphics[width=15cm,clip]{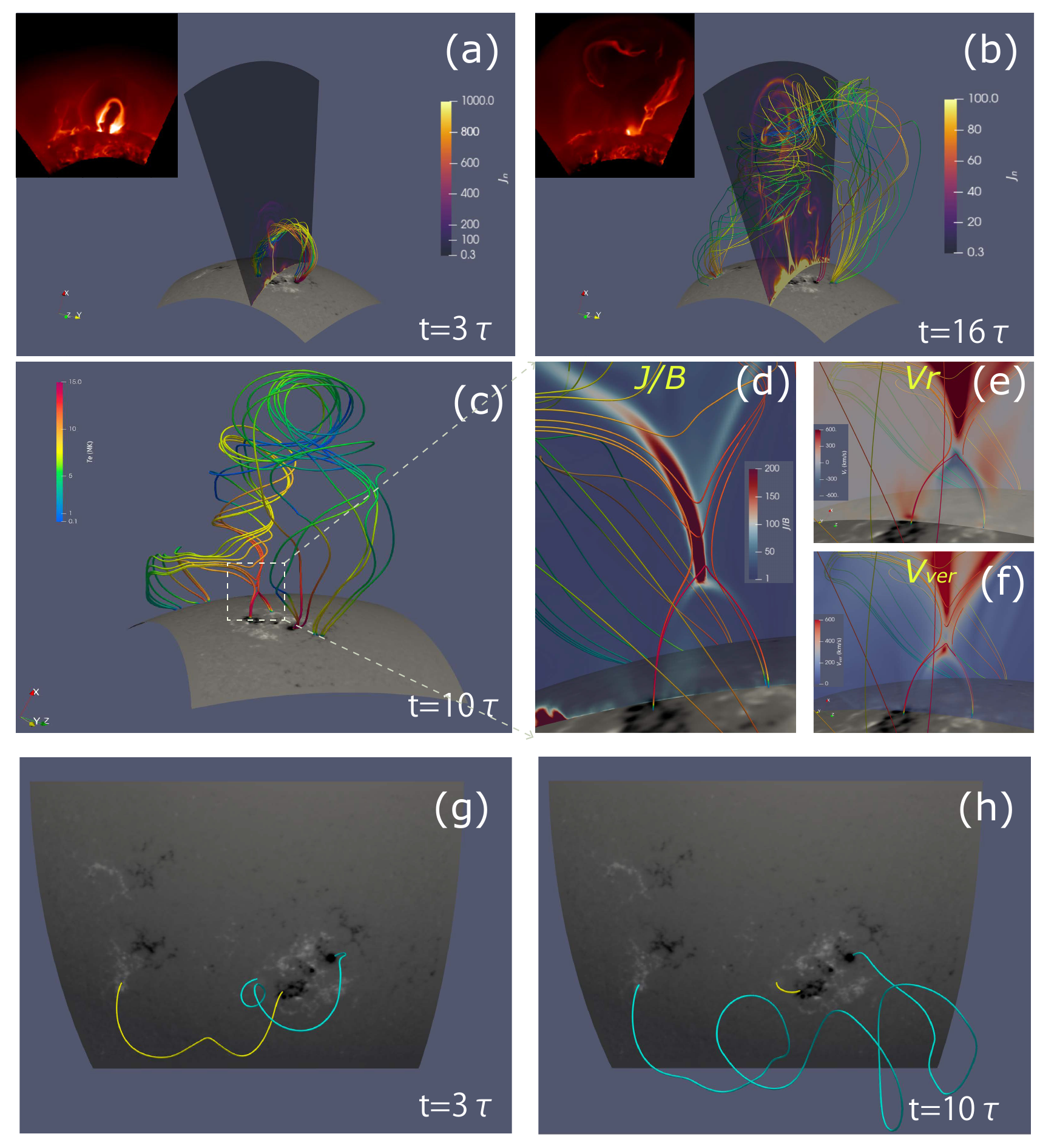}
  \centering
  \caption{Magnetic reconnection between eruptive flux rope and ambient magnetic field lines during the eruption. The field lines are colour-coded by temperature. Panels (a) and (b) illustrate the eruptive flux rope traced from the current intensity at 3 $\tau$ and 16 $\tau$, respectively, with the insets displaying the side-view synthesised radiation images in 304 \AA\ channel of SDO/AIA. Panel (c) depicts the reconnection configuration at 10 $\tau$. Panels (d)--(f) show the magnetic field lines along with the distribution of $J/B$, radial velocity $V_{r}$ and velocity perpendicular to the magnetic fields $V_{ver}$ near the current sheet, respectively. Panels (g) and (h) present two typical field lines traced from the same seed point in the photosphere at $t=$ 3 $\tau$ and 10 $\tau$, respectively. \label{figure6}}
\end{figure}

Figures~\ref{figure6}a and \ref{figure6}b display twisted flux-rope field lines traced from the circular electric current-carrying region and corresponding synthesised 304 \AA\ radiation images at two snapshots. At the early stage of the eruption ($t=$3 $\tau$), the field lines within the CME flux rope primarily wind around a common axis, anchored at two footpoints within the NOAA AR 12887, exhibiting an arch-like shape. This is in accord with the simulation result that the domain is restricted to the main eruptive active region \citep{Guo2023}. However, by $t=16 \ \tau$, certain twisted field lines extend to active region NOAA 12889, causing the CME flux rope footpoints to migrate. Besides, some cross-active-region twisted magnetic field lines are filled with cold plasma from the precursor filament. This suggests they may be formed through 3D magnetic reconnection between precursor flux-rope field lines and loops extending from neighbouring active regions. The synthesised images show the left drifting and extension to the nearby active regions.

To further examine the reconnection configuration, we present the 3D magnetic fields colour-coded by the temperature at $t=10\ \tau$ in Figure~\ref{figure6}c. A hot, X-shaped structure, composed of twisted flux-rope field lines and cross-active region loops, is visible, suggesting the occurrence of magnetic reconnection in which the flux rope takes part. To confirm this, we plot the distributions of $J/B$, radial velocity ($V_{r}$), and velocity perpendicular to the field lines ($V_{\rm ver}$) in Figures~\ref{figure6}d, \ref{figure6}e, and \ref{figure6}f, respectively. The X-shaped magnetic structure is surrounded by an elongated region of high $J/B$ values, indicating the presence of a current sheet conducive to magnetic reconnection. Moreover, bi-directional outflows resulting from this can be identified, with upflows driving the rise of CME flux rope and downflows impacting the underlying flare loops, causing them to shrink and oscillate in Figure~\ref{figure2}e. 

Furthermore, to investigate how the cross-active-region twisted flux rope is formed, we trace two typical field lines from the photosphere at $t=3\tau$ (Figure~\ref{figure6}g), including a cross-active-region loop (yellow line) and a twisted flux-rope line (cyan line). Figure~\ref{figure6}h shows the connectivity of these two field lines traced from the same seed points but at $t=10\ \tau$. The twisted flux-rope field line extends to the neighbouring active region 12889, while the cross-active-region loop transfers to a post-flare loop. As a result, we confirm that 3D magnetic reconnection across multiple active regions plays a crucial role in destroying the coherent structure of the CME flux rope.

\section{Conclusion and Discussion} \label{sec:sum}

This paper explores the birth of a major CME involving multiple active regions with imaging observations and a data-constrained MHD simulation. This work exhibits the impacts of 3D magnetic reconnection, driven by the presence of an overlying null-point structure between multiple active regions, on the magnetic-connectivity change of CME flux ropes and the produced phenomena in observations. This strongly suggests that the existence of multiple active regions is important in forming complex non-coherent CME flux ropes, departing from their progenitors. Moreover, according to our simulation results, we emphasize that the topology of CME flux ropes could be more complex than expected predicted in the simple bipolar configuration, especially for solar eruptions occurring at solar maximum.

\subsection{Origin of CMEs with intricate magnetic structure: 3D magnetic reconnection}

It is widely accepted that magnetic reconnection plays a pivotal role in solar eruptions. On the one hand, it is well accepted that magnetic reconnection can drive solar eruptions \citep{Chen2000, Karpen2012, Jiang2021, Jiang2024}. On the other hand, it can change the helicity of the CME flux rope, such as twist and toroidal flux \citep{Wangws2017, Jiang2021b, Guo2023b}. For example, in a 2D or 2.5D picture, magnetic reconnection in the enveloping magnetic fields of the CME flux rope can inject poloidal fluxes and cause its growth. Although the reconnection in the envelope magnetic field leads to forming a CME flux rope with multi-thermal distribution \citep{Zhao2017}, the coherent structure where field lines wind around one common axis is maintained in 2D scenarios. However, in a real 3D scenario, magnetic reconnection occurs at null points where $B=0$ and at separators and quasi-separators \citep{LiT2021}. Consequently, a greater variety of reconnection geometries is expected, forming CME flux ropes with more intricate topologies. 

Since magnetic flux ropes' connectivity significantly differs from the background magnetic fields, quasi-separatrix layers \citep[QSLs;][]{Priest1995} will naturally form to induce 3D magnetic reconnection involving flux ropes. For example, \citet{Aulanier2019} found that a flux rope can reconnect with ambient sheared arcades, characterised by an $ar-rf$ reconnection geometry (flux rope + arcade are reconnected to flux rope + flare loop). Hereafter, \citet{Guo2023b} investigated this reconnection geometry in an observational data-constrained MHD simulation and found that it also leads to the rotation of the CME flux rope and its departure from the progenitor filament. Besides, anti-parallel magnetic fields exist near the null point, indicating that the flux rope will likely reconnect with field lines extending from the null point during the passage. \citet{Jiang2013} explored an eruption below the null point and found that the eruptive flux rope reconnects with the null-point field lines. \citet{Driel-Gesztelyi2014} presented a filament eruption with drainage crossing multiple active regions and modelled the interaction between the CME flux rope and the ambient null point with a data-inspired model. This means that the 3D magnetic reconnection that flux ropes take part in can significantly alter the magnetic topology of CME flux ropes. Recently, \citet{Guo2024} simulated the propagation process of a CME from the solar surface to a distance of 25$R_{\odot}$ with a global coronal modelling coupling the solar wind. They found that the CME flux rope at around 20$R_{\odot}$ comprises open and closed twisted field lines with diverse footpoints, indicating that it should deviate from a coherent structure in the 2D picture. Here, based on observational data-constrained MHD simulations, we exhibit 3D reconnection geometry between eruptive flux rope and null-point field lines (Figure~\ref{figure4}), resulting in a CME flux rope with an intricate magnetic structure. First, it is seen that one leg or footpoint of the flux rope migrates to other active regions. Moreover, the non-coherent flux rope without one common axis is naturally formed. This strongly indicates that the magnetic structures of CMEs in real solar magnetic environments or observations should be more complicated than those obtained in simple bipolar configurations.

\subsection{The role of multiple active regions in the CME development}

In the solar-maximum decades, not only one active region may exist on the solar disk, thereby forming an overlying large-scale null point between active regions. As such, it is expected that the formation of a CME will probably involve multiple active regions. In such magnetic configuration, three types of reconnection geometries are expected: (1) reconnection in the background fields below the flux rope \citep{Jiang2021}; (2) reconnection at the null point above the flux rope \citep{Antiochos1999, Karpen2012}; and (3) reconnection at the interface between the flux rope and null-point fields \citep{Jiang2013, Driel-Gesztelyi2014}. Among them, the first two types, crucial in triggering solar eruptions, have been extensively in previous works \citep{Antiochos1999, Karpen2012, Jiang2021, Bian2023}. In this paper, we exhibit the critical role of third-type reconnection due to multiple active regions in altering the magnetic topology of CMEs, driven by multiple active regions.

\begin{figure}
  \includegraphics[width=18cm,clip]{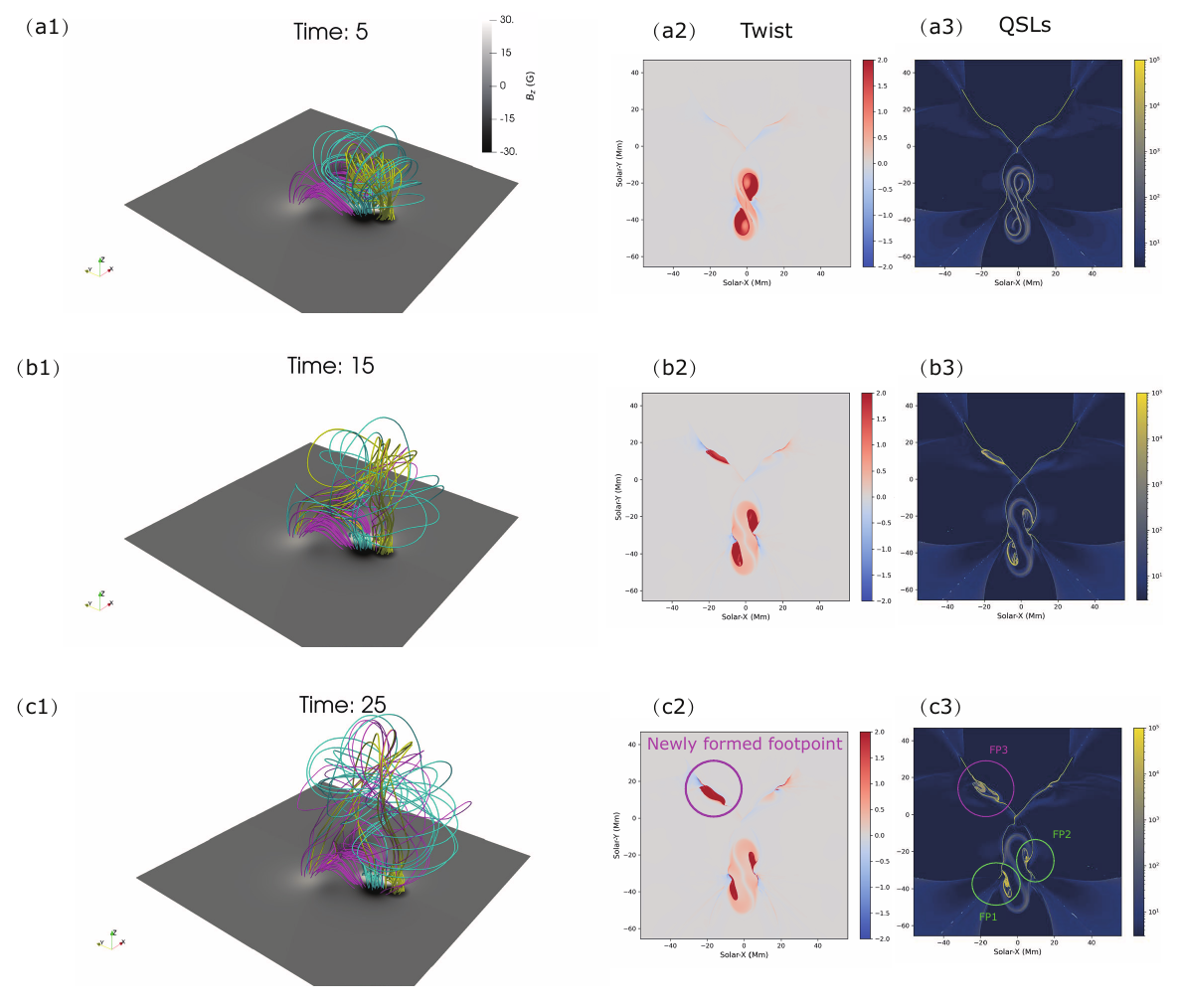}
  \centering
  \caption{Evolution of magnetic fields in the data-inspired modeling. The yellow, cyan, and wine-red field lines are traced from the pre-existing flux rope footpoint, the potential bipolar fields of the main active region, and the null-point magnetic fields between active regions, respectively. Panels from top to bottom correspond to times 5, 15, and 25 in dimensionless time. Panels (a1, b1, c1) show the 3D magnetic field configurations, (a2, b2, c2) depict the twist distributions on the bottom plane, and (a3, b3, c3) present the QSL distributions on the bottom plane. \label{figureS2}}
\end{figure}

To further validate the generality of the aforementioned physical process, we perform a data-inspired MHD simulation in a simple multipolar magnetic configuration (see Appendix~\ref{num2} for details on the numerical setup). Figure~\ref{figureS2} illustrates the evolution of the magnetic topology, including 3D magnetic field lines, twist, and QSL distributions on the bottom plane. Both data-constrained and data-inspired simulations exhibit a similar scenario: certain twisted flux-rope field lines connect to other active regions as the eruptive flux rope approaches the null point between active regions. In the twist and QSL distributions, a newly formed flux-rope footpoint (FP3), corresponding to the regions with $T_{w}>1$ and hook-shaped QSLs, is visible apart from the main eruptive active regions. Notably, the conjugate flux-rope footpoints in the main active region (FP1 and FP2) are nearly symmetrical at the start of the eruption (Figure~\ref{figureS2}a2). However, FP2 shrinks more significantly than FP1 with the appearance of FP3. In this scenario, it is crucial to consider the contributions of both bifurcated footpoints (FP2 and FP3) when quantifying the magnetic properties of CME flux ropes. As such, it is important to account for remote ribbons or dimmings when estimating magnetic fluxes of CMEs from their footpoints; otherwise, the fluxes of CMEs could be underestimated.

This simulation further confirms that a multiple active-region environment plays a crucial role in generating CMEs with intricate magnetic structures. In this scenario, a bifurcated CME flux rope with three footpoints, which deviates from a coherent structure, is naturally formed (Figure~\ref{figureS2}). This is consistent with the formation mechanism of pre-eruptive bifurcated flux ropes \citep{Zhong2019}. However, in simulations based on Titov-D\'emoulin-modified model \citep{Titov2014}, such as \citet{Liuqj2024a}, the twisted field lines predominantly wind around a common axis and are anchored at two footpoints in the photosphere. Furthermore, a comparison between \citet{Guo2023} and the data-constrained MHD simulation in this study supports a similar conclusion. In the simulation by \citet{Guo2023}, where the domain only covers NOAA AR 12887, the magnetic connectivity of eruptive flux rope is nearly aligned with its progenitor filament (though the western footpoint exhibits slight drift). In contrast, the CME flux rope in this paper, which incorporates an additional two active regions, deviates significantly from that in \citet{Guo2023}. These comparisons strongly underscore the significant impact of multiple active regions on the formation of complex, non-coherent CME flux ropes.

\subsection{Identifying magnetic reconnection involving magnetic flux ropes based on imaging observations}

Although the impacts of magnetic reconnection on the connectivity of CME flux ropes are challenging to identify with imaging observations due to the low density in the corona, there is indirect evidence for the occurrence of 3D magnetic reconnection based on reconnection productions (flare ribbons), connectivity change of CME flux ropes (filaments and hot channels) and their footprints in the low atmosphere (twin dimmings and the hook manifested by flare ribbons), which are detailed in the following paragraph. Identifying observational proxies for 3D magnetic reconnection is crucial for predicting the magnetic structures of CME flux ropes.

Previous works have suggested that the rooted positions and shapes of flare loops can be evidence for $ar-rf$ reconnection. \citet{Lorincik2019} found that the formation of flare loops rooted in the pre-eruptive flux rope footpoint when flare ribbons sweep them can be evidence of 3D magnetic reconnection. Hereafter, \citet{Lorincik2021} found that these flare loops commonly exhibit saddle-like shapes, in which the loops at cantles show higher and more inclined morphology. Additionally, changes in the topology of CME flux ropes manifested as eruptive filaments, can serve as direct evidence for the 3D magnetic reconnection that flux ropes take part in. For instance, \citet{Dudik2019} suggested that the shifts in filament legs represent reconnection between the eruptive flux rope and neighbouring loops. \citet{Driel-Gesztelyi2014} reported that certain filament materials drain to other active regions, indicating 3D magnetic reconnection between eruptive flux ropes and loops extending from other active regions. \citet{Guo2023b} found that lateral drifting of filament materials can serve as evidence of 3D magnetic reconnection, which leads to the result that the tilt angle of CME flux rope deviated from that of its progenitor filament, revealing the impacts of 3D magnetic reconnection on the axial orientation of CME flux ropes. In addition to coronal features like filaments, photospheric responses during eruptions, such as dimming and hook structure of flare ribbons, can also be used to identify 3D magnetic reconnection \citep{Sterling1997, Janvier2013}. The drifting of twin dimmings commonly represents 3D magnetic reconnection \citep{Aulanier2019}.

In this work, we verified that large-scale coronal jet crossing active regions can be used to determine 3D magnetic reconnection involving the interaction between flux ropes and multiple active regions. As shown in Figures~\ref{figure2}d and \ref{figure2}e, the coronal jet between active regions is visible when the eruptive filament passes through, which is rarely reported in previous studies. This physical process is also validated by the data-constrained simulation (Figure~\ref{figure6}) and the corresponding EUV synthesised radiation images (Figure~\ref{figure5}b). The following facts feature the jet in this paper: (1) it forms between active regions; (2) it connects two dimming regions; and (3) the back-flowing of filament materials along the jet spine accompanies it. These observational phenomena indicate that this jet involves interactions between the eruptive flux rope and multiple active regions. 

This work also highlights the connection between large-scale CMEs and small-scale solar activities. The jet observed between active regions is the large-scale counterpart to the small-scale blowout jet \citep{Moore2010}. In this scenario, the interchange reconnection between mini-filament and overlying open magnetic fields is commonly found \citep{Wyper2017, Li2023}. In our work, jets appear as the associated phenomena during CME expansion, representing 3D magnetic reconnection between the CME flux rope leg and ambient arcades. More importantly, this phenomenon also represents the birth of a CME with a complicated magnetic structure involving multiple active regions. In such cases, it is required to consider the impacts of nearby active regions on one CME when predicting its magnetic structure.

\begin{acknowledgments}

We thank the anonymous referee for detailed suggestions that helped to improve the quality of this paper, and the discussion with Y. K Kou, C. Xing, X. Cheng, Q. M. Zhang and W. T. Fu, P.F.C., Y.G. and J.H.G. is supported by the National Key R\&D Program of China 2020YFC2201200, 2022YFF0503004, NSFC (12127901) and China National Postdoctoral Program for Innovative Talents fellowship under Grant Number BX20240159. R.C.\ acknowledges the support from DST/SERB project No. EEQ/2023/000214. BS acknowledges support from KU Leuven. RJ acknowledges the support of the Research Council of Norway through its Centres of Excellence scheme, project number 262622. YZ acknowledges funding from the Research Foundation – Flanders FWO under project number 1256423N. The numerical calculations in this paper were performed in the cluster system of the High Performance Computing Center (HPCC) of Nanjing University. SP is funded by the European Union. However, the views and opinions expressed are those of the author(s) only and do not necessarily reflect those of the European Union or ERCEA. Neither the European Union nor the granting authority can be held responsible. This project (Open SESAME) has received funding under the Horizon Europe programme (ERC-AdG agreement No 101141362). These results were also obtained in the framework of the projects C16/24/010 C1 project Internal Funds KU Leuven), G0B5823N and G002523N (WEAVE) (FWO-Vlaanderen), and 4000145223 (SIDC Data Exploitation (SIDEX2), ESA Prodex).

\end{acknowledgments}

\bibliography{ms}{}
\bibliographystyle{aasjournal}

\section{appendix}
\label{appendix}

\subsection{Numerical setup of data-constrained modeling}\label{num}

To reveal physical details underlying the observations, we conduct a data-constrained, semi-relativistic, full thermodynamic MHD simulation for this eruptive event. Moreover, to incorporate multiple active regions, the simulation is established in spherical coordinates, which is more self-consistent in modelling the evolution of large-scale coronal structures than the models in Cartesian coordinates. Furthermore, to reproduce the thermodynamic evolution during the eruption, our modelling also considers the field-aligned thermal conduction, empirical heating and optically thin radiative cooling terms in the solar corona. The governing thermodynamic semi-relativistic MHD equations \citep{Gombosi2002} are as follows:

\begin{eqnarray}
 && \frac{\partial \rho}{\partial t} +\nabla \cdot(\rho \boldsymbol{v})=0,\label{eq1}\\
 && \frac{\partial }{\partial t}(\rho \boldsymbol{v}+\frac{1}{c^2 \mu_{0}}\boldsymbol{E} \times \boldsymbol{B})+\nabla \cdot(\rho \boldsymbol{vv}+p\boldsymbol{I}+\frac{B^2}{2\mu_{0}}\boldsymbol{I}-\frac{\ \boldsymbol{BB}}{\mu_{0}}+\frac{E^{2}}{2\mu_{0}c^2}\boldsymbol{I}-\frac{\boldsymbol{EE}}{\mu_{0}c^{2}}) \\
 \notag
 && =\frac{1}{\mu_{0}}(\frac{1}{c_{0}^2}-\frac{1}{c^2}) (\boldsymbol{E} \nabla \cdot \boldsymbol{E}-\boldsymbol{E}\times \nabla \times \boldsymbol{E})+\rho \boldsymbol{g},\label{eq2}\\
 && \frac{\partial \boldsymbol{B}}{\partial t} + \nabla \cdot(\boldsymbol{vB-Bv})=-\nabla \times (\eta \boldsymbol{J}),\label{eq3}\\
 && \frac{\partial }{\partial t}(\frac{p}{\gamma -1})+\nabla \cdot(\frac{p}{\gamma-1}\boldsymbol{v}) = - p \nabla \cdot
    \boldsymbol{v}+H - n_{\rm e}n_{\rm _{H}}\Lambda(T) + \nabla \cdot(\boldsymbol{\kappa} \cdot \nabla T)+Q, \label{eq4} 
\end{eqnarray}
where $\boldsymbol{E}=-\boldsymbol{v}\times \boldsymbol{B}$ is the motional electric-field vector, $\nabla \cdot(\boldsymbol{\kappa} \cdot \nabla T)$ is thermal conduction, $\kappa_{\parallel} =10^{-6}\ T^{\frac{5}{2}}\ \rm erg\ cm^{-1}\ s^{-1}\ K^{-1}$ is the Spitzer heat conductivity, $n_{\rm e}n_{\rm _{H}}\Lambda(T)$ is the optically-thin radiative losses. To maintain a high-temperature solar corona, we set an empirical background heating term ($H$) to balance the radiation losses, defined as $H=$ max $(H_{0}e^{-r/H_{l}}, H_{0}B^{1.75}n_{e}^{0.125}r^{-0.75})$, where $H_{0}=1.0\times10^{-6}\ \rm erg \ cm^{-3}\ s^{-1}$ represents the heating amplitude, $H_{l}=60$ Mm is the heating scale, $r$ is the local curvature radius of the field line. In addition, we also set a Joule heating term $Q = \eta J^2$, where $\eta=200\ \rm km^{2}\ s^{-1}$. To accurately control the magnetic-field divergence during the numerical computation, we adopt the constrained transport (CT) method in the staggered grid \citep{Gardiner2005}, such that the magnetic-field divergence can be maintained to machine precision. Especially, the original magnetograms are directly adopted in this model without any reduction \citep{Gombosi2002}, meaning that the magnetic fields in our MHD model can reach a value of $\sim$ 2500 G in observations. As a result, both the evolution of thermodynamics and magnetic fields are more comparable to real observations. 

For the boundary conditions, the horizontal electric fields ($E_{\rm th}$ and $E_{\rm ph}$) in the bottom cell are set to zero, ensuring that the $B_{\rm r}$ component remains constant. The velocities are set to zero in the ghost cells of all six boundaries. The magnetic fields are specified using equal-gradient extrapolation at the bottom boundary and zero-gradient extrapolation at the other five side boundaries. For density and pressure, they are fixed at their initial values at the bottom boundary, flexible at the top with the hydrostatic equilibrium condition, and provided with equivalent extrapolation at the four side boundaries. As a result, the bottom cell surface satisfies the line-tied condition. The partial differential equations in spherical coordinates are numerically solved with the Message Passing Interface Adaptive Mesh Refinement Versatile Advection Code \citep[MPI-AMRVAC\footnote{http://amrvac.org},][]{Xia2018, Keppens2023}. The computational domain is $[r_{min},r_{max}]\ \times [\theta_{min},\theta_{max}]\ \times [\phi_{min},\phi_{max}] = [1\ R_{\odot}, 2\ R_{\odot}]\ \times [92.6^{\circ},131.4^{\circ}]\ \times[-37.4^{\circ},25.6^{\circ}]$, with an effective grid of $384 \times 384 \times 384$, using a four-level adaptive mesh refinement with a radially stretched grid. Therefore, the effective spatial resolution in $r, \theta, \phi$ can reach around 40 km, $0.1^{\circ}$ and $0.164^{\circ}$.

Similar to \citet{Guo2024}, this modelling can be separated by relaxation and eruption stages. The relaxation stage is used to obtain the solar atmosphere coupled with magnetic fields, wherein the initial magnetic field is provided by the potential field constructed with the method of \citet{Fisher2020}. The density and pressure are given by a hydrostatic atmosphere, including the chromosphere and the solar corona. It should be noted that the eruptive flux rope is not included in this stage to realise a fast relaxation. To achieve thermodynamic equilibrium of the atmosphere in conjunction with magnetic fields, we relax the above state until the mean pressure and velocity in the computation domain remain nearly constant. 

Hereafter, we insert a twisted magnetic flux rope with the regularised Biot-Savart Laws (RBSL) method \citep{Titov2018} to drive the solar eruption. The implementation process in MPI-AMRVAC can be seen in \citet{Guoy2023}. The reconstruction of an RBSL flux rope requires four parameters: toroidal flux ($F$), electric current ($I$), cross-section radius ($a$) and flux-rope path ($C$). After conducting several numerical tests, we set the toroidal flux ($F$) to $10^{20}$ Mx, cross-section radius ($a$) to 45 Mm, and the path is outlined by the pre-eruptive filament in observations. As such, the electric current flowing through the flux rope can be derived directing in the RBSL method \citep{Titov2018}. Additionally, to be more consistent with observations, we insert a filament by increasing the density by two orders of magnitude while maintaining constant pressure at magnetic dips of the flux rope. It should be pointed out that the inserted flux rope is at the height of about 45 Mm, which is higher than the critical height of torus instability ($\sim$ 25 Mm) and eruptive height ($\sim$ 15 Mm) according to the analysis of \citet{Duan2023}. Additionally, the current sheet is also seen below the flux rope at such height so that the fast reconnection can also result in the eruption based on the scenario of \citet{Jiang2021}. This means that the inserted flux rope is not stable and will erupt once switched to the eruption stage. Figure~\ref{figure3} displays the initial magnetic fields and the inserted dense filament at the start of the eruption stage, defined as $t=0$.

\subsection{Radiation synthesis}\label{radia}

To directly compare simulation results with observations, we perform radiation synthesis in EUV wave channels of SDO/AIA using the Radiation Synthesis Tools \footnote{https://github.com/fuwentai/radsyn\_tools}. With the assumption of an optically thin model, the radiation emission in each computation cell is calculated with the following formula:

\begin{eqnarray}
I_{\lambda}(r, \theta, \phi)=G_{\lambda}(T)n_{e}^2(r, \theta,\phi)
\end{eqnarray}
where $I_{\lambda}$ represents the radiation emission, $G_{\lambda}(T)$ is the response function for different wavebands, and $n_{e}$ denotes the number density. Hereafter, the radiation image can be computed by integrating the emission along the line of sight.

\subsection{Numerical setup of data-inspired modeling in simple multipolar configuration}\label{num2}

Here, we introduce the numerical setup of the data-inspired MHD simulation in a multipolar magnetic configuration. In this model, the core active region responsible for the eruption is constructed using TDm model \citep{Titov2014}. Two sub-photospheric magnetic charges ($q=50$) are placed at a depth of 1 and separated by a normalized distance of 2 to provide strapping fields. Additionally, two groups of potential bipolar fields are introduced to mimic adjacent active regions, defined by the following formula \citep{Aulanier2010}:

\begin{eqnarray}
B_{x}=\sum_{i=1}^{4} C_{i}(x-x_{i})r_{i}^{-3}\\
B_{y}=\sum_{i=1}^{4} C_{i}(y-y_{i})r_{i}^{-3}\\
B_{z}=\sum_{i=1}^{4} C_{i}(z-z_{i})r_{i}^{-3}\\
\end{eqnarray}
where ($x_{1}=-4$;\ $y_{1}=6$;\ $z_{1}=-2$;\ $C_{1}=50$), ($x_{2}=-4$;\ $y_{2}=4$;\ $z_{2}=-2$;\ $C_{2}=-30$), ($x_{3}=4$;\ $y_{3}=6$;\ $z_{3}=-2$;\ $C_{3}=-50$), ($x_{4}=4$;\ $y_{4}=4$;\ $z_{4}=-2$;\ $C_{4}=30$). The flux rope is positioned at a height of 20 Mm, with an electric current intensity that is 1.3 times the equilibrium current calculated from Shafranov’s equilibrium condition. This ensures the flux rope is not fully balanced, leading to its immediate eruption. As such, our data-inspired model closely replicates the observed magnetic field distributions, including the main eruptive active region containing the eruptive flux rope (filament) and two surrounding active regions. Consistent with observations, an overlying null point between the active regions is also found.

Since this model mainly aims to demonstrate the role of multiple active regions in forming CMEs with complex magnetic structures, we employ the zero-$\beta$ MHD model to simulate the evolution of the CME flux rope, as described below:

\begin{eqnarray}
 && \frac{\partial \rho}{\partial t} +\nabla \cdot(\rho \boldsymbol{v})=0,\label{eq1}\\
 && \frac{\partial (\rho \boldsymbol{v})}{\partial t}+\nabla \cdot(\rho \boldsymbol{vv}-\boldsymbol{BB})+\nabla (\frac{\boldsymbol{B}^2}{2})=0,\label{eq2}\\
 && \frac{\partial \boldsymbol{B}}{\partial t} + \nabla \cdot(\boldsymbol{vB-Bv})=0,\label{eq3}
\end{eqnarray}
where $\rho$, $\boldsymbol{v}$ and $\boldsymbol{B}$ represent the density, velocity, and magnetic field, respectively. The initial density is the same as in \citet{Guo2019}, and the velocity is set to zero throughout the domain. The computational domain is defined as ${[x_{min},x_{max}]\times[y_{min},y_{max}]\times[z_{min},z_{max}]=[-150,150]\times[-100,200]\times[3,303]}$ Mm$^{3}$, with a uniform grid resolution of 160 $\times$ 200 $\times$ 200. It should be noted that, unlike the thermodynamic data-constrained MHD simulation presented in this paper, this zero-$\beta$ model omits the effects of gravity and thermal pressure. Nevertheless, the 3D evolution of the magnetic fields remains physically reasonable due to the low plasma $\beta$ environment in the solar atmosphere. The squashing degree $Q$ \citep{demo1996, Titov2002} and twist number is computed with the open-source code of \citet{Liu2016}.

\end{document}